\documentclass[twocolumn,prd,showpacs,superscriptaddress,groupedaddress,preprintnumbers,nofootinbib]{revtex4-1}
\usepackage{graphics}
\usepackage{graphicx,amsfonts,amsmath,amssymb,epsfig,color, mathrsfs,latexsym,url,wasysym,float,slashed,dsfont}
\newcommand{\be}{\begin{equation}}
\newcommand{\ee}{\end{equation}}

\newcommand{\ba}{\begin{eqnarray}}
\newcommand{\ea}{\end{eqnarray}}

\begin{document}
\preprint{RCHEP/23-002}
\title{On Torsion Contribution to Chiral Anomaly via Nieh-Yan Term}
\author{Ida M. Rasulian$^1$}
\author{Mahdi Torabian$^2$}
\affiliation{$^1$School of Physics, Institute for Research in Fundamental Sciences (IPM), P.O.Box 19395-5531, Tehran, Iran\\
$^2$Research Center for High Energy Physics, Department of Physics, Sharif University of Technology\\Azadi Ave., P.Code 1458889694, Tehran, Iran}
\begin{abstract}
In this note we present a solution to the question of whether or not, in the presence of torsion, the topological Nieh-Yan term contributes to chiral anomaly. The integral of Nieh-Yan term is non-zero if topology is non-trivial; the manifold has a boundary or vierbeins have singularities. 
Noting that singular Nieh-Yan term could be written as a sum of delta functions, we argue that the heat kernel expansion cannot end at finite steps. This leads to a sinusoidal dependence on the Nieh-Yan term and the UV cut-off of the theory (or alternatively the  minimum length of spacetime). We show this ill-behaved dependence can be removed if a quantization condition on length scales is applied. It is expected as the Nieh-Yan term can be derived as the difference of two Chern class integrals ({\it i.e.} Pontryagin terms). On the other hand, in the presence of a  cosmological constant, we find that indeed the Nieh-Yan term contributes to the index with a dimensionful anomaly coefficient that depends on the de Sitter length or equivalently inverse Hubble rate. We find similar result in thermal field theory where the anomaly coefficient depends on temperature. In both examples, the anomaly coefficient depends on IR cut-off of the theory. Without singularities, the Nieh-Yan term can be smoothly rotated away, does not contribute to topological structure and consequently does not contribute to chiral anomaly.
\end{abstract}
\maketitle
\section{Introduction}
\label{intro}
There has been a long debate with conflicting results on the contribution of torsion to chiral anomaly via the Nieh-Yan term \cite{Shapiro,Zanelli1,Zanelli2,Mielke1,Mielke2,Maroto2,Ubukhov,Goldthorpe,Li,Peeters}. One suspects that in the presence of torsion, the parity odd Nieh-Yan four-form could potentially result in non-conservation of chiral current\footnote{We recall that in the presence of torsion, the Pontrygin term constructed from curvature is modified and includes axial part of torsion \cite{Alvarez}. In this note, we study the contribution from the Nieh-Yan term.}. However, as the vierbein is chosen dimensionless and torsion has mass dimension one, the anomaly coefficient of the Nieh-Yan term is dimensionful. Naive computations imply that the mass parameter in the coefficient is the UV cut-off of the theory, namely the anomaly term is regulator dependent which is not normally accepted. It is the source of confusion that led some authors to completely abandon the Nieh-Yan part. 

In this short paper, we revisit this problem and we show that depending on the underlying spacetime, the Nieh-Yan term may or may not contribute to chiral anomaly. In fact for smooth geometry, the Nieh-Yan contribution is completely rotated away. On the other hand, in geometries with singularities in vierbein structure, it needs more elaboration. In this case, in a locally flat gauge condition, the Nieh-Yan term can be written as a sum of delta functions about singular points. We will argue that, in the Fujikawa method with a regulator, the Heat kernel expansion cannot be truncated at finite steps. It gives a sinusoidal dependence on the Nieh-Yan term as well as on the UV cut-off. This ill contribution can be removed via applying a quantization condition on the parameters. We argue that it is natural as the Nieh-Yan term can be written as a difference between two Pontryagin terms. However, we find that in the presence of cosmological constant, there is a residual and indeed the Nieh-Yan term contributes to the index and chiral anomaly. The anomaly coefficient is proportional to the square of the Hubble rate or inverse de Sitter radius. Similarly, we find that in a finite temperature field theory there is a contribution to chiral anomaly through the Nieh-Yan term. In this case, the anomaly coefficient is proportional to temperature squared. In both scenarios, the anomaly is related to a mass scale, in fact the IR scale in the theory. 

In the next section, after a brief review of chiral anomaly and torsional geometry we present the main results and finally conclude in the section III.
\section{Torsion and Chiral Anomaly}
\label{sec:1}
We briefly review the calculation of chiral anomaly using the Fujikawa method. In this method one uses the invariance of partition function under a chiral rotation, which is merely a field redefinition, and uses the fact that the measure in field space does not remain invariant under this redefinition. Considering an infinitesimal chiral rotation given by parameter $\beta(x)$ as
\ba\label{transf}
\psi'(x)=e^{i\beta(x)\gamma^5}\psi(x), \\ \bar{\psi}'(x)=\bar{\psi}(x)e^{i\beta(x)\gamma^5}.
\ea
the Dirac Lagrangian transforms as
\be 
L'=\bar{\psi}' i\slashed{D} \psi' = L-\partial^\mu \beta \bar{\psi}\gamma_\mu \gamma^5 \psi.
\ee
The partition function can be written as
\be 
S=\int d\psi d\bar{\psi} \; exp\Big[\int d^4 x L\Big].
\ee
Under the chiral transformation \eqref{transf} the measure $d\psi d\bar{\psi}$ is not left invariant and transforms as
\be 
d\psi' d \bar{\psi}'=J(\beta) d\psi d \bar{\psi},
\ee  
where we have 
\be 
J(\beta)=exp\Big[-2i \int d^4x \beta(x)\sum_n \psi_n^\dag \gamma^5 \psi_n\Big],
\ee
and $\psi_n$ are the eigenmodes of the Dirac operator.

Invariance of partition function under this transfromation further implies
\be\label{noncons}
\partial_\mu J^{\mu 5}=\sum_n \psi_n^\dag \gamma^5 \psi_n,
\ee
where $J^{\mu 5}=\bar{\psi}\gamma^\mu \gamma^5 \psi$ is the chiral current.

In general this sum is not well defined as it stands since this is equal to $tr \gamma^5. \delta^4(0)$, therefore it is common to use a regulator to make it well defined. The typical regulator that one can use is $exp\big[-\frac{\lambda_n^2}{M^2}\big]$ where $\lambda_n$ are the eigenvalues of the Dirac operator and $M$ is some UV cut-off that is sent to infinity in the calculation.

On the other hand we can see that the integral of non-conservation of chiral current is determined by a topological quantity, namely the index of the Dirac operator $i\slashed{D}$ on a manifold ${\cal M}$. Considering the integral of \eqref{noncons} we see that only the zero modes contribute to the integral and we can write 
\be 
ind = \int d^4 x \partial_\mu J^{\mu 5}=n_+-n_-,
\ee
where $n_+$ and $n_-$ are respectively the number of positive and negative chirality zero modes.

Using this background we can write the index of the Dirac operator as
\ba\label{indexdef} ind &=& \int_{\cal M} {\rm d}^4 x   \lim_{M\rightarrow\infty}\lim_{x'\rightarrow x}\sum_n\psi_n^{\dagger}(x')\gamma^5e^{-\frac{\slashed{D}^2}{M^2}} \psi_n(x) \cr &=& -\frac{1}{2}\int_{\cal M}{\rm d}(\star j_5),\ea
where $\psi_n$ are the eigenstates of the Dirac operator and we have used the Fujikawa method with a Gaussian cut-off fixed by $M$ to regularize the sum. 
Anomaly computations imply that the chiral current non-conservation is proportional to parity odd Pontryagin and Nieh-Yan \cite{Alvarez,Nieh1,Nieh2,Nieh3} four forms
\be {\rm d}(\star j_5) =\frac{1}{24\pi^2} (F^{ab}\wedge F_{ab}) + \frac{M^2}{4\pi^2}(T^a\wedge T_a - R_{ab}\wedge e^a\wedge e^b),\ee
where the field strengths (curvature and torsion respectively) are 
\be F_{ab}= D{\rm A_{ab}} = {\rm d}{\rm A_{ab}} +{\rm A_{ac}}\wedge {\rm A^c_b},\quad T^a = De^a = {\rm d}e^a+{\rm A^a_b}\wedge e^b,\ee
which satisfy Bianchi identities $DF_{ab} = 0$ and $DT_a = F_{ab}\wedge e^b$. Here $e^a=e^a_\mu dx^\mu$ is the vierbein and $\rm A$ is a connection 1-form.

We note the appearance of a dimensionful coefficient $M$ in the second term which is also expected on dimensional reasons as the vierbein is dimensionless. If we decompose the Lorentz connection into spin-connection plus contorsion {\it i.e.} ${\rm A_{ab}}=\omega_{ab}+K_{ab}$, then the curvature is written as $F_{ab}= R_{ab}+DK_{ab}+K_{ac}\wedge K^c_b$ where $R_{ab}=d\omega_{ab}+\omega_{ac}\wedge\omega^c_b$ is Riemann curvature and torsion would be $T_a = K_{ab}\wedge e^b$. 
Then, the Pontryagin term is computed as
\be F^{ab}\wedge F_{ab} = 
 R^{ab}\wedge R_{ab} - \frac{1}{4}{\rm d}S\wedge {\rm d}S, \ee
where the axial vector field $S$ is the fully antisymmetric part of torsion ({\it a.k.a.} H-torsion)
\be S = \star (e^a\wedge T_a),\ee 
or in component form $S_\mu = \epsilon_{\mu\nu\rho\sigma}T^{\nu\rho\sigma}$. This is the part of torsion that 
couples to fermions through the Dirac operator\cite{Nascimento} $\slashed{D}=\slashed{D}_0-\frac{i}{8} \slashed{S}\gamma^5$ 
($\slashed{D}_0$ is the torsion-free operator). Moreover, the axial vector $S$ can be decomposed into a transverse pseudo-vector $S_\mu^\perp$ (namely $\partial_\mu S^{\perp \mu} =0$) and a pseudo-scalar $\sigma$ 
\be S_\mu=S_\mu^{\perp}+\partial_\mu \sigma.\ee
Finally, the Nieh-Yah term which only receives a contribution from the pseudo-scalar part of torsion is  \cite{Li} 
\be
T^a\wedge T_a-R_{ab}\wedge e^a \wedge e^b = {\rm d}\,(e^a\wedge T_a) = {\rm d}(\star S) = -\frac{1}{4}{\bf 1}\square \sigma,
\ee
where ${\bf 1}$ is the unit (volume) four-form. 


On the other hand, the Nieh-Yan term can be written as the difference of two Chern class integrals \cite{Nieh3}\cite{Zanelli2}. We consider two SO(1,4) connections  
\be\label{connections}
{\rm A}^{AB}=\begin{pmatrix}\omega^{a b} & \frac{1}{l}e^a \\ -\frac{1}{l}e^b & 0 \end{pmatrix}\qquad {\rm and}\qquad 
{\rm A}^{AB}_0=\begin{pmatrix}\omega^{a b} & 0 \\ 0 & 0 \end{pmatrix},
\ee
where $l$ is a length scale. Then, the field strengths are
\be
F^{AB}=\begin{pmatrix}R^{a b}-\frac{1}{l^2}e^a\wedge e^b & \frac{1}{l}T^a \\ -\frac{1}{l}T^b & 0 \end{pmatrix}\ {\rm and}\   
F^{AB}_0=\begin{pmatrix}R^{a b} & 0 \\ 0 & 0 \end{pmatrix}.
\ee
We find that the difference between corresponding Chern class integrals is proportional to the Nieh-Yan term
\be\label{diff} 
F^{ab}\wedge F_{ab}-F_{0}^{ab}\wedge F_{0 ab}=\frac{2}{l^2}(T^a\wedge T_a-R_{ab}\wedge e^a \wedge e^b).
\ee
Since these are topological terms with integer integrals, this leads to a quantization condition on the Nieh-Yan term that has a dependence on the length scale $l$. We think a natural choice for this length scale where we want to have dimensionless geometry \cite{Volovik1,Volovik2} is the Planck's length considered to be the minimum length of spacetime.

\subsection{Manifolds with Trivial Topology}
\label{sec:2}
In this section we consider the contribution of non-singular Nieh-Yan term to chiral anomaly. Pseudo-scalar $\sigma$ has a non-singular behavior and using the redefinition $\psi(x)=e^{\frac{i}{8} \sigma(x)\gamma^5}\phi(x)$ we can see that the index does not depend on the total derivative part of axial torsion. In particular, we find
\be \mathcal{L}= i\bar{\psi}(x)\slashed{D}\psi(x)= i\bar{\phi}(x)\slashed{D}_0\phi(x).
\ee
By considering the Ward identity
\be \langle\partial_{\mu}J^{\mu 5}(x)\rangle=\frac{\int \mathcal{D}\psi(x)\mathcal{D}\bar{\psi}(x) e^{\int d^4x' i\bar{\psi}(x')\slashed{D}\psi(x')}\partial_{\mu}J^{\mu 5}(x') }{\int \mathcal{D}\psi(x)\mathcal{D}\bar{\psi}(x) e^{\int d^4x' i\bar{\psi}(x')\slashed{D}\psi(x')}},
\ee
we note that the field redefinition is like a chiral transformation the measure in numerator and denominator is multiplied by the same factor thus we find
\be \langle\partial_{\mu}J^{\mu 5}(x)\rangle_\psi=\langle\partial_{\mu}J^{\mu 5}(x)\rangle_\phi.
\ee
We conclude the total derivative part of axial torsion doesn't contribute to chiral anomaly.

Moreover, if we consider the Gaussian regulator $e^{-\frac{\slashed{D}^2}{M^2}}$ in Fujikawa method with
\ba \slashed{D}^2=\slashed{D}^{2}_0 &+& i\nabla^{\mu}S^\perp_{\mu}\gamma^5+S^{\perp2} \cr &-& 2e^{\mu}_a e^{\nu}_b \sigma^{ab}S^\perp_{\mu}D^0_{\nu}\gamma^5+ e^{\mu}_a e^{\nu}_b \sigma^{ab}\nabla_{\mu}S^\perp_{\nu}\gamma^5,\quad
\ea 
where $\slashed{D}$ only has the $S^\perp_{\mu}$ part, we find that the contribution from torsion to chiral anomaly is
\ba \nabla_\mu J^{\mu 5} &\supset& a \Box \nabla_{\mu}S^{\perp\mu}+b \nabla_{\mu}(S^{\perp2} S^{\perp\mu}) \cr &+& c \nabla_{\mu}(R^{\mu\nu}S^\perp_{\nu}-\frac{1}{2}R S^{\perp\mu})+{\rm d}S^\perp\wedge {\rm d}S^\perp.
\ea
Using $\nabla_{\mu}S^{\perp\mu}=0$ and the Bianchi identity 
\be \nabla_{\mu}(R^{\mu\nu}-\frac{1}{2}R g^{\mu\nu})=0,
\ee
we find
\ba \nabla_\mu J^{\mu 5} &\supset&b S^{\perp\mu} S^{\perp\nu} \nabla_{\mu}S^\perp_{\nu}+c R^{\mu \nu}\nabla_{\mu}S^\perp_{\nu}+{\rm d}S\wedge {\rm d}S.\quad
\ea
Apparently, the first two terms depend on our choice of $\sigma$ and $S^\perp_{\mu}$. There is a residual symmetry transformation $S^\perp_{\mu}\rightarrow S^\perp_\mu-\partial_\mu C, \sigma\rightarrow \sigma+C$ with $\Box C=0$, thus we can choose a gauge such that the sum of first two terms vanishes. Therefore, the chiral anomaly would be 
\be \langle\nabla_{\mu}J^{\mu 5}\rangle= R^{ab}\wedge R_{ab}+{\rm d}S\wedge {\rm d}S.
\ee

We provide an explanation for why calculating anomaly with the typical regulator $e^{-\frac{\slashed{D}^2}{M^2}}$ does not lead to the expected result. If we define a transformation
\be
\psi'=e^{i \beta(x)\gamma^5 e^{-\frac{\slashed{D}^2}{M^2}}}\psi,
\ee
\be
\bar{\psi}'=\bar{\psi}e^{i \beta(x)\gamma^5 e^{-\frac{\slashed{D}^2}{M^2}}},
\ee
using the partition function we can show that the anomaly of this transformation is
\be
\langle\partial_{\mu}(\bar{\psi}\gamma^{\mu}\gamma^5 e^{-\frac{\slashed{D}^2}{M^2}}\psi)\rangle=\sum_n \psi_n^\dagger \gamma^5 e^{-\frac{\lambda_n^2}{M^2}} \psi_n.
\ee
So we see that using the regulator is equivalent to calculating the anomalous current $\langle\partial_{\mu}(\bar{\psi}\gamma^{\mu}\gamma^5 e^{-\frac{\slashed{D}^2}{M^2}}\psi)\rangle$. We can show that this current is not the same as the original current by looking at a symmetry that is present in the original calculation which is missing here. The transformation
\be\psi \rightarrow e^{i \sigma(x)\gamma^5}\psi, S_\mu \rightarrow S_\mu-\partial_\mu \sigma,
\ee
leaves $\langle\partial_{\mu}(\bar{\psi}\gamma^{\mu}\gamma^5 \psi)\rangle$ invariant but the above current is not left invariant under this transformation. 

We note that even without the above explanation, we can redefine the chiral current such that the total derivative contributions like $\nabla_\mu K^\mu=\nabla_\mu (S^2 S^\mu)+...$ are removed from the anomaly without changing its topological structure. This is also true for the contribution of the non-singular part of the Nieh Yan term as this part behaves like a total derivative and is not a topological term.

\subsection{Manifolds with Non-Trivial Topology}
We begin this section by an observation that on a manifold with singularities in its vierbeins, the Nieh-Yan term can be written as a sum of delta functions \cite{Li} 
\be\label{ny}
T^a\wedge T_a-R_{ab}\wedge e^a \wedge e^b=-\frac{1}{4}{\bf 1}\square \sigma={\bf 1}\sum_i l_i^2 \delta^4(x-x_i),
\ee
where $l_i$ are some length scales and $x_i$ are singular points. In the following, for the sake of simplicity, we ignore spacetime curvature and only consider the contribution of torsion to anomaly via the Nieh-Yan term. In the Fujikawa method, the regulator has an extra piece due to the Nieh-Yan term
\be \slashed{D}^2=\slashed{D}^{2}_0-\frac{i}{8}\square \sigma\gamma^5+\frac{1}{64}\partial_\mu \sigma \partial^\mu \sigma+\cdots.
\ee
Then, we find the torsion contribution to the index as
\ba ind \supset \lim_{M\rightarrow \infty}\int \frac{{\rm d}^4k}{(2\pi)^4}e^{-\frac{k^2}{M^2}}{\rm tr}\Big[\gamma_5e^{i\frac{\square\sigma}{8M^2}\gamma_5}\Big].\ea
Given the delta function form of Nieh-Yan term, we cannot stop the sum at the first order term in $\square \sigma$ in the expansion and we must consider all terms. Then, the torsion contribution to the index is computed as 
\ba 
ind &\supset & i\lim_{M\rightarrow\infty}\frac{M^4}{16\pi^2}\int {\rm d}^4 x\, {\rm tr}\big[\gamma^5 e^{\frac{i\square \sigma}{8M^2}\gamma^5}\big] \cr \!\!&=& -\lim_{M\rightarrow\infty}\frac{M^4}{4\pi^2}\int {\rm d}^4 x\sin\Big(\frac{-\textstyle{\sum_i}l_i^2\delta^4(x-x_i)}{2M^2}\Big).\quad\ 
\ea
In order to calculate this term we use a simple method. Using the expansion of sine function we write
\begin{multline}
M^4 \int d^4 x \sin\Big( \frac{l_i^2 \delta^4(x)}{2M^2}\Big)\\=M^4 \int d^4 x \sum_{n=1}^\infty \frac{(-1)^n}{(2n+1)!} \Big(\frac{l_i^2 \delta^4(x)}{2M^2}\Big)^{2n+1}\\=M^4 \int d^4 x \Big(\frac{l_i^2 \delta^4(x)}{2M^2}\Big)\sum_{n=1}^\infty \frac{(-1)^n}{(2n+1)!} \Big(\frac{l_i^2 \delta^4(x)}{2M^2}\Big)^{2n}.
\end{multline}
Using the definition of delta function $\int d^4 x \delta^4(x)f(x)=f(0)$ and the approximation $\delta^4(0)=\frac{1}{\epsilon^4}$ with $\epsilon$ related to the minimum length of spacetime (and therefore related to the UV cut-off) we can write
\ba
ind \supset \frac{M^4 \epsilon^4}{4\pi^2}\sin\Big(\frac{l_i^2}{2M^2 \epsilon^4}\Big),
\ea
and considering $M=\frac{k}{\epsilon}$ with $k$ some order 1 number we find
\ba
ind \supset \frac{k^4}{4\pi^2}\sin(\frac{l_i^2}{2k^2 \epsilon^2}).
\ea
Thus, in order to remove this term in the index, we demand a quantization condition as
\be\label{quant}
\frac{l_i^2}{2M^2 \epsilon^4}=n_i \pi,
\ee
where $n_i$ are integer numbers.
We note that, if we had incorrectly kept only the first order term in the expansion of $e^{i\frac{\square \sigma}{8 M^2}}$, then we would have found a contribution to the index as 
\be 
ind \supset \frac{M^2 l_i^2}{4\pi^2},
\ee
which would be cut-off dependent and is not accepted. However, if we keep all terms in the expansion we find
\be \label{ind-1}
ind \supset \frac{M^4 \epsilon^4}{4\pi^2}\sin\Big(\frac{l_i^2}{2M^2 \epsilon^4}\Big),
\ee
which is finite given that $\epsilon\sim \frac{1}{M}$ and vanishing if the quantization condition \eqref{quant} is met.

Same result can be obtained via another approach. We recall that the zero modes of the Dirac operator with torsion $\psi^0_n$ and the zero modes of the Dirac operator without torsion $\phi^0_n$ are related as
\be\label{redef}
\psi^0_n(x)=e^{\frac{i}{8} \sigma(x)\gamma^5}\phi^0_n(x).
\ee
With a finite torsion term, this simple reasoning states that the total derivative part of pseudo-vector torsion does not contribute to the index. However with singular point, the outcome is different. Given \eqref{ny} we write
\be 
\sigma(x)=-4\sum_i \frac{l_i^2}{|x-x_i|^2},
\ee
as we have singular behavior at some points. Taking care of the divergent behavior close to one of the singularities we find
\ba
ind\!&\supset&\!\!\int\!\! {\rm d}^4 x  \lim_{x'\rightarrow x}\sum_n {\rm tr}\Big[\gamma^5e^{i\frac{\sigma(x)}{8}\gamma^5 } \phi^0_n(x)\phi_n^{0\dagger}(x')e^{-i\frac{\sigma(x')}{8}\gamma^5}\Big]\cr 
&=&\!\!\int\!\! {\rm d}^4 x  \lim_{x'\rightarrow x}\sum_n{\rm tr}\Big[\gamma^5 e^{-i\frac{\sigma(x')-\sigma(x)}{8}\gamma^5} \phi^0_n(x)\phi_n^{0\dagger}(x')\Big].\ \ \ \ \
\ea
Applying  
\ba
e^{-i\frac{\sigma(x')-\sigma(x)}{8}\gamma^5}&=&\cos\Big(\frac{1}{8} (x'-x)^\mu\partial_\mu \sigma\Big)\cr &-&i \gamma^5 \sin\Big(\frac{1}{8}(x'-x)^\mu\partial_\mu \sigma \Big),
\ea 
the index is computed as 
\ba\label{index}
ind\! &\supset&\!\! \int\!\! {\rm d}^4 x  \lim_{x'\rightarrow x}\cos\!\Big(\frac{1}{8}(x'-x)^\mu\partial_\mu \sigma \Big)\!\sum_n {\rm tr}[\gamma^5 \phi^0_n(x)\phi_n^{0\dagger}(x')]
\cr &&\qquad\quad\ -i\sin\Big(\frac{1}{8} (x'-x)^\mu\partial_\mu\sigma\Big)\delta^4(x-x').
\ea
Given a minimum length scale $\epsilon$ we may write 
\be 
\frac{1}{8}(x'-x)^\mu\partial_\mu \sigma  \sim\frac{l_i^2}{2\epsilon^2}.
\ee
The first term in \eqref{index} is proportional to the torsion-free Index. The second part is divergent though. Again, if we demand a quantization condition for $l_i$ as
\be
\frac{l_i^2}{2\epsilon^2}\sim n_i \pi,
\ee 
where $n_i$ are integers, then the divergent piece will be absent. 

In passing, we emphasis that this quantization condition is expected as the Nieh-Yan term could be seen as the difference between two Chern classes \cite{Nieh3,Zanelli1} where here we replace the length scale in \eqref{diff} with the minimum length of spacetime. One might therefore conclude that adding torsion to gravity has a close relationship with spacetime discretization. In particular, if we choose $l$ to be the minimum length of spacetime, we find the quantization condition
\be\label{quantum}
\frac{l_i^2}{\epsilon^2}=4\pi^2 n_i.
\ee
Therefore, if in \eqref{quant} we choose $M=\frac{\sqrt{2\pi}}{\epsilon}$ the sinusoidal dependence of index to the Nieh-Yan term will be absent.

In the following we consider two cases both with a natural IR scale and show that conterary to what previously expected the coefficient of the Nieh-Yan term will be related to the IR scale of the theory. In this context we use the IR cut-off as the lower limit for the physical energy levels. In the case of a positive cosmological constant this natural scale is given by the cosmological constant. This can be expected if we consider de Sitter spacetime as a compact Euclidean space with effective radius $l$. In the presence of temperature we also have a natural IR scale which is expected given the fermionic Matsubara frequencies $\omega_n=(2n+1)\pi T$.
\subsubsection*{1. Adding a Cosmological Constant}
In the absence of the cosmological constant, as the Nieh-Yan term could be seen as the difference of two Pontryagin terms, we considered two Lorentz connections in \eqref{connections}
and derive 
\be 
\frac{l_i^2}{l^2}=4\pi^2 n_i,
\ee
where $n_i$ is an integer and we have chosen $\epsilon=l$.

In the presence of the cosmological constant $\Lambda \sim l_0^{-2}$ with de Sitter length $l_0$, we propose to change the second Lorentz connection in \eqref{connections} as
\be\label{l0}
{\rm A}^{AB}_0=\begin{pmatrix}\omega^{a b} & \frac{1}{l_0}e^a \\ -\frac{1}{l_0}e^b & 0 \end{pmatrix}.
\ee
Then, the field strengths are
\be
F^{AB}=\begin{pmatrix}R^{a b}-\frac{1}{l^2}e^a\wedge e^b & \frac{1}{l}T^a \\ -\frac{1}{l}T^b & 0 \end{pmatrix},
\ee
\be
F^{AB}_0=\begin{pmatrix}R^{a b}-\frac{1}{l_0^2}e^a\wedge e^b & \frac{1}{l_0}T^a \\ -\frac{1}{l_0}T^b & 0 \end{pmatrix}.
\ee
We find that the difference between corresponding Chern class integrals is proportional to the Nieh-Yan term as
\be\label{diffr} 
F^{ab}\wedge F_{ab}-F_{0}^{ab}\wedge F_{0 ab}=2\Big(\frac{1}{l^2}-\frac{1}{l_0^2}\Big)(T^a\wedge T_a-R_{ab}\wedge e^a \wedge e^b).
\ee
Then, following the same line of reasoning we find the quantization condition
\be\label{quanta}
\Big(\frac{1}{l^2}-\frac{1}{l_0^2}\Big)l_i^2=4\pi^2 n_i.
\ee
In the previous section, we computed the torsion contribution to anomaly 
\be 
ind \supset \sin \Big(\frac{l_i^2}{2M^2 l^4}\Big).
\ee
Upon using \eqref{quanta} we find
\be 
ind\supset\sin\Big(\pi n_i+\frac{l_i^2}{4\pi l_0^2}\Big)=(-1)^{n_i} \sin\Big(\frac{l_i^2}{4\pi l_0^2}\Big)\approx (-1)^{n_i} \frac{l_i^2}{4\pi l_0^2},
\ee
which is similar to the result suggested by \cite{Zanelli2}.
Therefore, we see from \eqref{ind-1} that the coefficient of Nieh-Yan term in index should be the cosmological constant as an IR cut-off of the theory and we can write
\be 
ind \supset (-1)^{n_i}\Lambda_{IR}^2 (T^a\wedge T_a-R_{ab}\wedge e^a\wedge e^b).
\ee

Note that the appearance of a potentially non-integer index is a result of spacetime quantization and is evident from \eqref{indexdef}. When we take the limit $x'\rightarrow x$ there is always a mismatch due to discreteness of spacetime. On the other hand due to the quantization condition \eqref{quanta} the final result for the index is equivalent to replacing the UV cut-off in the calculation with the IR scale $\frac{1}{l_0}$. This is equivalent to summing over the eigenmodes of the Dirac operator with eigenvalues less than the IR scale in the index {\it i.e.} these modes effectively play the role of zero-modes in a theory  with an IR scale.

\subsubsection*{2. Finite Temperature Effect}
Finally, we study the effect of finite temperature on chiral anomaly in a background with torsion. It is in particular interesting in condensed matter experiments where torsion is realized as dislocations or defects in a sample \cite{Kondo,Katanaev,Kleinert}. 
In the presence of temperature we use the concept of periodic Euclidean time in the definition of delta function. We have
\be\label{sum}
\delta_\beta (t-t_i)=\sum_n \delta(t+n \beta-t_i),
\ee
where $\beta=\frac{1}{T}$ is inverse temperature. We can approximate the delta function as
\be 
\delta(t)=\frac{1}{\pi}\frac{\epsilon}{\epsilon^2+t^2},
\ee
where $\epsilon$ is related to the minimum length of spacetime. Then, after calculating the sum in \eqref{sum} we find
\ba 
\delta_\beta (t)&=&\frac{i}{2\beta}\Big(\cot \frac{\pi(t+i\epsilon)}{\beta}-\cot \frac{\pi(t-i\epsilon)}{\beta}\Big)\cr&\approx& \frac{1}{\pi}\frac{\epsilon}{\epsilon^2+t^2}+\frac{\pi \epsilon}{3\beta^2},
\ea
which is plotted in Fig.1. Then, we find
\be\label{temp}
\delta_\beta(t-t_i)=\frac{\epsilon}{\pi(\epsilon^2+(t-t_i)^2)}+\frac{\pi \epsilon}{3\beta^2}=\delta(t-t_i)+\frac{\pi \epsilon}{3\beta^2}.
\ee
\begin{figure}
\begin{center}
\includegraphics[scale=.5]{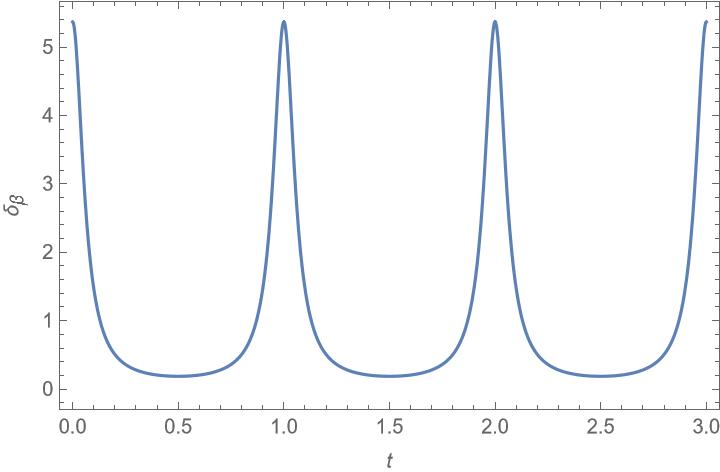}
\end{center}
\caption{The approximated Dirac delta function with periodic argument for $\epsilon=0.06$ and $\beta=1$.}
\label{fig:1}       
\end{figure}

The torsion contribution to the index is computed as
\be 
ind\supset \sin(\sum_i \frac{l_i^2 \delta_\beta^4(x-x_i)}{2M^2}).
\ee
Using \eqref{temp} and $M^2=\frac{2\pi}{\epsilon^2}$, we find 
\be 
ind\supset\sum_i (-1)^{n_i}\frac{T^2}{12}l_i^2.
\ee 
It is consistent with the results presented in \cite{Volovik,Stone} if we require $n_i$ to be odd.

\section{Conclusions}
In this short note, we argued that there could exist an extra torsional contribution to chiral anomaly through Nieh-Yan term if the manifold has singular points and the field theory has a natural IR cut-off. Moreover for that to be a sound result, we found that there must exist a quantization condition on the minimum length scale in the present of torsion. We studied two scenarios which come with an IR cut-off; one with a cosmological constant and one in a finite temperature. In both cases we found a finite contribution to chiral anomaly proportional to the IR cut-off. These scenarios are in particular interesting as they are respectively relevant in early Universe cosmology and condensed matter experiments. The latter has been already studied in literature (see for instance \cite{Volovik,Stone}). In a future work, we will study the effect of this new torsional contribution to chiral current anomaly during inflation. 
\medskip
\paragraph*{Acknowledgements} 
This work is supported by the research deputy of Sharif University of Technology. We are thankful to Hessamaddin Arfaei for fruitful discussions. 

\section*{Appendix A}
In this appendix, we argue that we can subtract the effect of non-topological terms like $\nabla_\mu (S^2 S^\mu)$, $\nabla_\mu(\square S^\mu)$ and the non-delta function part of the Nieh-Yan term $\nabla_\mu S^\mu$ by redefining the current or equivalently adding counter terms to the Lagrangian. These counter terms include $S^4$, $S^\mu \square S_\mu$ and $M^2 S^2$ which correspond to redefining the chiral current by adding terms proportional to $S^2 S^\mu$, $\square S^\mu$ and $M^2 S^\mu$. This subtraction is allowed as far as the topological behaviour of chiral anomaly is not altered.

We explain how we can subtract the effect of the extra terms using these counter terms. Considering the Lagrangian 
\be 
L=i\bar{\psi}(\slashed{D}+i\slashed{S}\gamma^5)\psi+ a S^4 + b S^\mu \square S_\mu + c M^2 S^2,
\ee
where $a,b,c$ are some numbers and variating with respect to $S_\mu$ we find
\be 
J^{\mu 5}= a S^2 S^\mu +b \square S^\mu+ c M^2 S^\mu. 
\ee
Therefore we see that by an appropriate choice of $a,b,c$ we can subtract the effect of the mentioned finite terms from chiral anomaly.

In passing we note that the presence of a coupling $S_\mu C^\mu$ in the effective Lagrangian in \cite{Nascimento}, where $C^\mu$ is related to Pontryagin as $P=2\nabla_\mu C^\mu$, is expected since we can write $J^{\mu 5}=\frac{\delta L_{eff}}{\delta S_\mu}$ and the $C^\mu$ term in this variation is a confirmation of the presence of Pontryagin term in chiral anomaly.

In the case of delta-function type Nieh-Yan term we can write
\be
\nabla_\mu S^\mu= \sum_i l_i^2 \delta^4(x-x_i).
\ee
As an example we consider the case of a sphere as a compact manifold with two such terms one at the north pole and one at the south pole say $x_1$ and $x_2$. We need to cover the sphere with two patches each including one of these singularities. We can write
\be 
\int \nabla_\mu S^\mu =\int_+ S_+^\mu {\rm d}a_\mu -\int_- S_-^\mu {\rm d}a_\mu,
\ee
where $S_+$ and $S_-$ are respectively defined in patch 1 and patch 2. So the integral depends on the difference of $S$ defined in the intersection of the two patches. Now given that $S^\mu \approx \frac{l_i^2 (x^\mu-x_i^\mu)}{|x-x_i|^4}$ and considering the behaviour of $d a_\mu$ as we take the limit $x\rightarrow \infty$ (that we assume to be the boundary of each patch) we find as expected that this integral receives a non-zero contribution from delta-function type Nieh Yan term.

Using the same argument we can show that terms like $\int S^2 S^\mu {\rm d} a_\mu$ and $\int \square S^\mu {\rm d} a_\mu$ receive no contributions from delta-function type Nieh-Yan term and therefore can be subtracted from anomaly by redefining the chiral current.

We show that this reasoning is in fact a result of our previous argument that a total derivative axial torsion should not contribute to anomaly as far as the topological behaviour is not altered. Our original Lagrangian is 
\be\label{lag}
L=i \bar{\psi}\slashed{\partial} \psi - S_\mu \bar{\psi} \gamma^\mu \gamma^5 \psi +k M_{Pl}^2 S^\mu S_\mu.
\ee
Here the $S^2$ term appears from torsional Ricci scalar. We have seen that when we use the regulator $e^{-\frac{\slashed{D}^2}{M^2}}$ we are in fact calculating the divergence of a different current $\bar{\psi} \gamma^\mu \gamma^5 e^{-\frac{\slashed{D}^2}{M^2}} \psi$. This is effectively equivalent to changing the Lagrangian to 
\be
L=i \bar{\psi}\slashed{\partial} \psi - S_\mu \bar{\psi} \gamma^\mu \gamma^5 e^{-\frac{\slashed{D}^2}{M^2}} \psi +k M_{Pl}^2 S^\mu S_\mu.
\ee
Expanding to first order we find extra terms like
\be\frac{1}{M^2} S_\mu \bar{\psi}\gamma^\mu\gamma^5(\square+S^2+...)\psi.\ee On the other hand if we solve for $S_\mu$ in \eqref{lag} classically we find 
\be 
J^{\mu 5} = 2k M_{Pl}^2 S^\mu.
\ee
Therefore, if we require $M\propto M_{Pl}$ to be regarded as the UV cut-off, the additional terms considering this identity will be $S^\mu \square S_\mu$ and $S^4$ among other terms. This means that if we use the regulator $e^{-\frac{\slashed{D}^2}{M^2}}$ we need to subtract the effect of these extra terms which is equivalent to redefining the chiral current.

\end{document}